\documentclass[prd,aps,a4paper,superscriptaddress,twocolumn,nofootinbib]{revtex4}
\usepackage{graphicx}
\usepackage{color}
\usepackage{xcolor}
\usepackage{dcolumn}
\usepackage{bm}
\usepackage{slashed}
\usepackage{amsmath}
\usepackage{latexsym}
\usepackage{amssymb}
\usepackage{mathrsfs}
\usepackage{amsfonts}
\usepackage{url}
\usepackage{hyperref}
\usepackage{xspace}
\allowdisplaybreaks

\newcommand{\NRSurMemory}{\texttt{NRSurMemory\_7qd4}\xspace}

\begin{document}
\title{Gravitational wave memory of the binary black hole events in GWTC-2}

\author{Zhi-Chao Zhao}
\affiliation{Department of Astronomy, Beijing Normal University,
Beijing 100875, China}
\author{Xiaolin Liu}
\affiliation{Department of Astronomy, Beijing Normal University,
Beijing 100875, China}
\author{Zhoujian Cao
\footnote{corresponding author}} \email[Zhoujian Cao: ]{zjcao@amt.ac.cn}
\affiliation{Department of Astronomy, Beijing Normal University,
Beijing 100875, China}
\affiliation{School of Fundamental Physics and Mathematical Sciences, Hangzhou Institute for Advanced Study, UCAS, Hangzhou 310024, China}
\author{Xiaokai He}
\affiliation{School of Mathematics and
Computational Science, Hunan First Normal University, Changsha
410205, China}

\begin{abstract}
Gravitational wave (GW) memory is an important prediction of general relativity. Existing works on the GW memory detection focus on the waveform analysis. It is hard for waveform analysis method to detect the GW memory due to its quasi-direct current behavior and weakness. We implement a completely different scheme in this work to estimate the GW memory. In this scheme, we firstly apply the Bondi-Metzner-Sachs method to calculate the GW memory of binary black hole based on numerical relativity simulation. Then we construct a surrogate model to relate binary black hole's parameters and the GW memory. Afterwards we apply this surrogate model together with Bayesian techniques to estimate the GW memory of the 48 binary black hole events recorded in GWTC-2. The GW memory corresponding to the all 48 events has been estimated. The most interesting results are for GW190814. The corresponding GW memory is about $-1\times10^{-23}$ and $1\times10^{-23}$ for Hanford detector and Livingston detector respectively. At the same time we find with 3$\sigma$ C.L. that the memory strain of GW190814 is negative on Hanford detector while positive on Livingston detector.
\end{abstract}

\maketitle

\section{Introduction}
The memory of gravitational wave (GW) was firstly found by Zeldovich, Braginsky, Thorne and their coworkers \cite{Zeldovich74,Pay83,Braginsky:1986ia,braginsky1987gravitational}. This kind of GW memory is produced by the change of a quadrupole moment for slowly moving sources. Christodoulou found that gravitational wave itself can also produce memory \cite{PhysRevLett.67.1486,Fra92}. This kind of memory is usually called nonlinear memory. The GW memory detection \cite{PhysRevLett.121.071102,PhysRevD.101.023011,PhysRevD.101.083026} may be used to study the gravitational theory \cite{PhysRevD.94.104063} and spacetime dimension \cite{Hollands_2017}. Several works in the past years \cite{Set09,VanLev10,PshBasPos10,CorJen12,MadCorCha14,Arzoumanian_2015,PhysRevLett.117.061102,PhysRevLett.118.181103,PhysRevD.102.023010} have investigated the possibility of detecting the nonlinear memory. All of the works focused on the waveform analysis. Because the GW memory behaves mainly as a quasi-direct current signal, detector responses to it weakly. Such fact makes GW memory detection hard.

In this paper, we implement an alternative method to investigate the GW memory of the binary black hole merger events recorded by LIGO and VIRGO. Firstly we design a Bondi-Metzner-Sachs method to calculate the GW memory based on numerical relativity simulation of binary black holes. Then we apply this method to the SXS catalogue \cite{SXSBBH} and construct a database of binary black hole intrinsic parameters and the corresponding GW memory due to the gravitational radiation. Based on such a database we use Gaussian process regression to construct a surrogate model describing the relationship between the binary black hole's parameters and the gravitational wave memory. Afterwards we apply such surrogate model together with Bayesian techniques to infer the GW memory of the 48 binary black hole (BBH) events in GWTC-2. We can well estimate the GW memory of these BBH events by the use of parameters' samples based on the analysis of GW waveforms. And our analysis indicates that our estimation admits high confidence.

Our work is similar but different to that of \cite{PhysRevD.103.044012}. Ref.~\cite{PhysRevD.103.044012} only studied the spherical harmonic mode $h_{20}$ while we investigate the detector's expected response to the GW memory.
Mode $h_{20}$ only depends on GW source's intrinsic parameters while the detector's response depends also on extrinsic parameters. But only the detector's response is observable and measurable for GW experiment.

In the next section we describe the Bondi-Metzner-Sachs (BMS) method of GW memory calculation used in the current work \cite{PhysRevD.103.043005}. Then we combine the BMS method and numerical relativity simulations to construct a surrogate model of GW memory for BBHs. After that we apply our surrogate model to GW memory estimation of the binary black hole events in GWTC-2 of LIGO. Finally we give a summary and a discussion.

\section{Method of gravitational wave memory calculation}
The Newmann-Penrose components of Weyl tensor $\Psi_\mu,\mu=0,...,4$ admit the following relation in the wave zone \cite{BonVanMet62,Sac62,PenRin88}
\begin{align}
\dot{\Psi}_2=\eth\Psi_3+\sigma\Psi_4,\,\, \Psi_3=-\eth\dot{\bar{\sigma}},\,\, \Psi_4=-\ddot{\bar{\sigma}}.\label{eq1}
\end{align}
Here $\sigma$ corresponds to the shear of the $(\theta,\phi)$ coordinate sphere in the Bondi-Sachs coordinate \cite{he2015new,he2016asymptotical,sun2019binary}. The overbar means complex conjugate. The $\eth$ operator is related to the sphere geometry. The overdot means the time derivative. The shear $\sigma$ is related to the gravitational wave strain through
\begin{align}
\sigma=\frac{D}{2}\left(h_++ih_\times\right),
\end{align}
where $D$ is the luminosity distance between the observer and the source, $h_{+}$ and $h_{\times}$ correspond to the two polarization modes of the gravitational wave.
The relations (\ref{eq1}) result in
\begin{align}
\frac{\partial}{\partial t}(\Psi_2+\bar{\sigma}\dot{\sigma})
=|\dot{\sigma}|^2-\eth^2\dot{\bar{\sigma}}
+\bar{\sigma}\ddot{\sigma}-\sigma\ddot{\bar{\sigma}}.\label{meq1}
\end{align}
We can use spin-weighted $-2$ spherical harmonic functions to decompose the gravitational wave strain $h\equiv h_{+}-ih_{\times}$ as following \cite{PhysRevD.75.124018,PhysRevD.77.024027,PhysRevD.78.124011}
\begin{align}
&h(t,\theta,\phi)\equiv\sum_{l=2}^{\infty}\sum_{m=-l}^lh_{lm}(t)Y_{-2lm}(\theta,\phi),
\end{align}
where $Y_{slm}$ means spin-weighted $s$ spherical harmonic function. Plug the above decomposition into Eq.~(\ref{meq1}) we get
\begin{align}
&h_{lm}\bigg{|}_{-\infty}^{+\infty}=-\sqrt{\frac{(l-2)!}{(l+2)!}}\left[\left.\frac{4}{D}\int \Psi_2Y_{0l0}\sin\theta d\theta d\phi\right|_{-\infty}^{+\infty}\right.-\nonumber\\
&D\sum_{l'=2}^{\infty}\sum_{l''=2}^{\infty}\sum_{m'=-l'}^{l'}\sum_{m''=-l''}^{l''}
\Gamma_{l'l''lm'-m''0}\times\nonumber\\
&\,\,\,\,\,\left.\left(\int_{-\infty}^{+\infty}\dot{h}_{l'm'}\dot{\bar{h}}_{l''m''}dt-\dot{h}_{l'm'}\bar{h}_{l''m''}\bigg{|}_{-\infty}^{+\infty}\right)\right].\label{meq2}\\
&\Gamma_{l'l''lm'-m''-m}\equiv\nonumber\\
&\,\,\,\,\,\,\,\,\,\,\,\,\int Y_{-2l'm'}\overline{Y}_{-2l''m''}\cdot\overline{Y}_{0lm}\sin\theta d\theta d\phi.
\end{align}
Now we decompose $h_{lm}=h^{\rm osc}_{lm}+h^{\rm mem}_{lm}$ where $h^{\rm osc}_{lm}$ is the oscillation part which means $h^{\rm osc}_{lm}(-\infty)=h^{\rm osc}_{lm}(+\infty)=0$ and $h^{\rm mem}_{lm}$ is the memory part which means $\dot{h}^{\rm mem}_{lm}\approx0$ due to the quasi-direct current (DC) behavior of the GW memory \cite{Fav09a}. Then the above Eq.~(\ref{meq2}) becomes
\begin{align}
&h^{\rm mem}_{lm}\bigg{|}_{-\infty}^{+\infty}=-\sqrt{\frac{(l-2)!}{(l+2)!}}\left[\left.\frac{4}{D}\int \Psi_2Y_{0l0}\sin\theta d\theta d\phi\right|_{-\infty}^{+\infty}\right.-\nonumber\\
&D\sum_{l'=2}^{\infty}\sum_{l''=2}^{\infty}\sum_{m'=-l'}^{l'}\sum_{m''=-l''}^{l''}
\Gamma_{l'l''lm'-m''0}\times\nonumber\\
&\,\,\,\,\,\left.\int_{-\infty}^{+\infty}\dot{h}^{\rm osc}_{l'm'}\dot{\bar{h}}^{\rm osc}_{l''m''}dt\right].\label{meq4}
\end{align}

If we take the mass center frame of the BBH system at the past infinity time as the inertial frame, we have $\Psi_2(-\infty,\theta,\phi)=M$. Here $M$ corresponds to the BBH's initial total mass (Bondi mass) \cite{Ashtekar:2019viz}. At the future infinity time, the BBH's total mass $M'=M-E_{\rm GW}$ measured in the above inertial frame is smaller than the initial value $M$ because the gravitational wave carries away some energy $E_{\rm GW}$. The spacetime will settle down to a Kerr black hole with mass $\tilde{M}$ at the future infinity time. But importantly the mass center frame at the future infinity time is different to the above inertial frame corresponding to the mass center frame at the past infinity time due to the kick velocity. These two inertial frames corresponding to the mass center frame at past infinity time and the mass center frame at the future infinity time are related by a boost transformation described by the kick velocity. Consequently $\tilde{M}=M'/\gamma$, where $\gamma$ is the Lorentz factor. So corresponding to the Eq.~(\ref{meq2}) we have \cite{Ashtekar:2019viz}
\begin{align}
&\Psi_2(+\infty,\theta,\phi)=-\frac{\tilde{M}}{\gamma^3}\times\nonumber\\
&\left(1-v_x\sin\theta\cos\phi-v_y\sin\theta\sin\phi-v_z\cos\theta\right)^{-3},\label{meq3}\\
&\gamma=\frac{1}{\sqrt{1-v^2}},
\end{align}
where $v_x$, $v_y$ and $v_z$ are the Cartesian components of the kick velocity $v$.

Since the gravitational wave energy $E_{\rm GW}$, the kick velocity and the oscillation part $h^{\rm osc}_{lm}$ have already been accurately obtained by numerical relativity simulation \cite{PhysRevD.103.024031}, we can plug them into the Eq.~(\ref{meq4}) to calculate the gravitational memory $\left.h^{\rm mem}_{lm}\right|_{-\infty}^{+\infty}$. In Ref.~\cite{PhysRevD.103.043005} we investigated the waveform of GW memory. Differently here we care about the overall GW memory of BBH coalescence.

\begin{figure}
\begin{tabular}{c}
\includegraphics[width=0.47\textwidth]{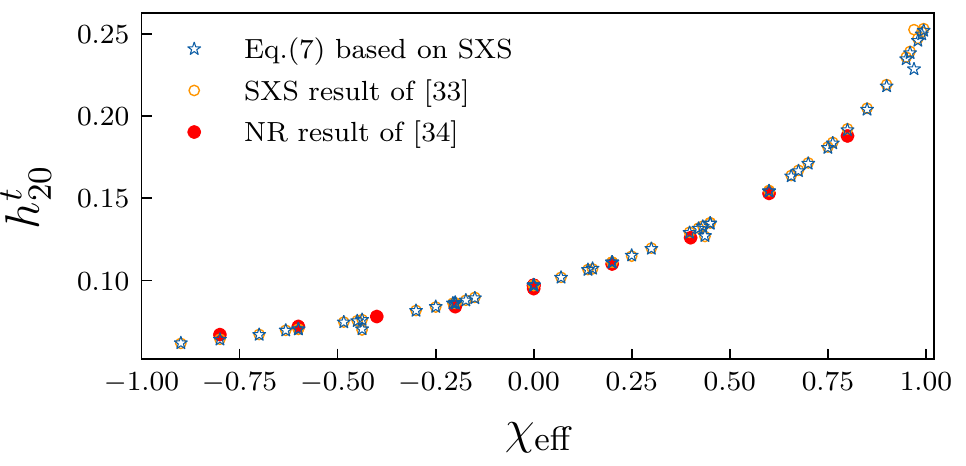}
\end{tabular}
\caption{GW memory $h^t_{20}$ of spin aligned equal mass BBH respect to the effective spin. The NR result of GW memory is borrowed from the Table.~1 of \cite{PolRei11}.}\label{fig1}
\end{figure}

In Fig.~\ref{fig1} we compare our calculation results
\begin{align}
&h^t_{lm}\equiv\left.\frac{D}{M}h^{\rm mem}_{lm}\right|_{-\infty}^{+\infty}
\end{align}
for spin aligned equal mass BBH systems based on SXS simulations to the numerical relativity results by direct calculation in \cite{PolRei11}. 
Here $h^t_{lm}$ refers the ``intrinsic'' memory and it is not an observable quantity. However, after multiplying the factor $\frac{D}{M}$ and 
projecting to a detector, it will be observable(see Sec. III).
In addition the recent memory calculation results by SXS group \cite{PhysRevD.103.024031} are also shown in the figure for comparison. Following the convention of \cite{PolRei11} we use the effective spin $\chi_{\rm eff}\equiv\frac{s_{1z} m_1+s_{2z} m_2}{m_1+m_2}$ as the horizontal axis
, where $s_{1z,2z}$ and $m_{1,2}$ denotes the spins' $z$-component and component masses of the two black holes. The perfect consistency indicates the reliability of our method for GW memory calculation. For these spin aligned BBHs, we confirm previous approximation that $h^{\rm mem}_{lm}\approx0,m\neq0$ and $h^{\rm osc}_{l0}\approx0$. For precession BBHs, our results got through the Bondi-Metzner-Sachs method are also consistent to that of SXS results \cite{PhysRevD.103.024031} like Fig.~\ref{fig1}.

\section{Surrogate model of gravitational wave memory for BBH}
\begin{figure}
\includegraphics[width=0.47\textwidth]{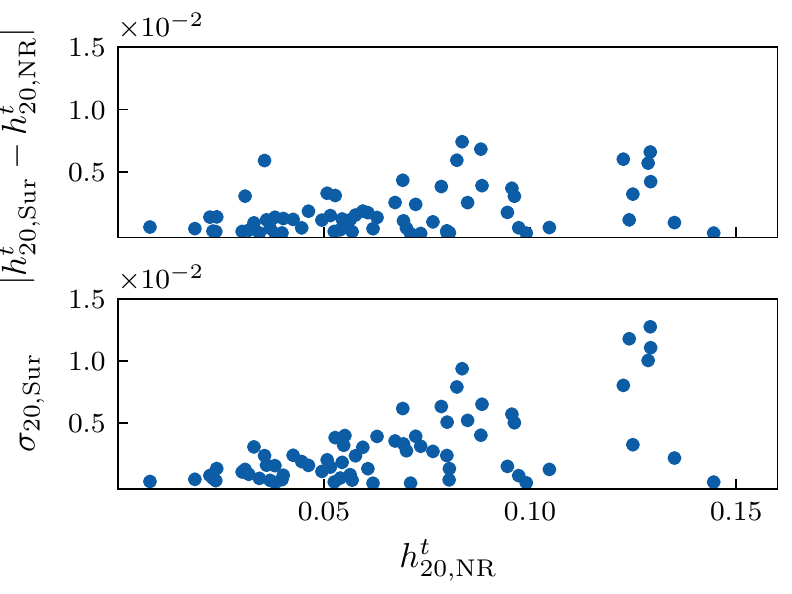}
\caption{Accuracy of the surrogate model \NRSurMemory against the 69 test samples. Top: The difference between the model prediction $h^t_{\rm 20Sur}$ and the direct calculation result $h^t_{\rm 20NR}$ through the Bondi-Metzner-Sachs method based on the numerical relativity simulations.  Bottom: The estimated error by the \NRSurMemory model.}\label{fig2}
\end{figure}

\begin{figure}
    \includegraphics[width=0.47\textwidth]{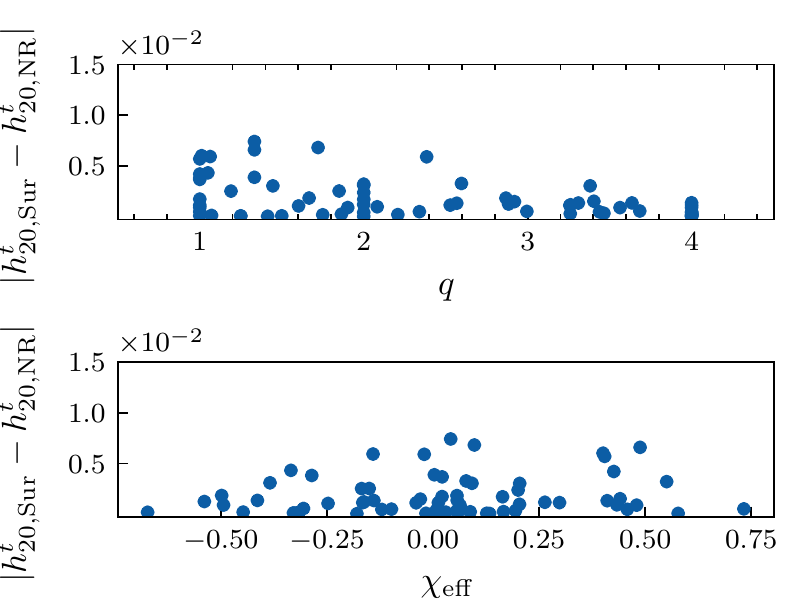}
    \caption{The dependence of the errors in Fig.~\ref{fig2} with mass ratio $q$(Top) and effective spin $\chi_{\rm eff}$(Bottom). }\label{fig2A}
\end{figure}

\begin{figure}
    \includegraphics[width=0.47\textwidth]{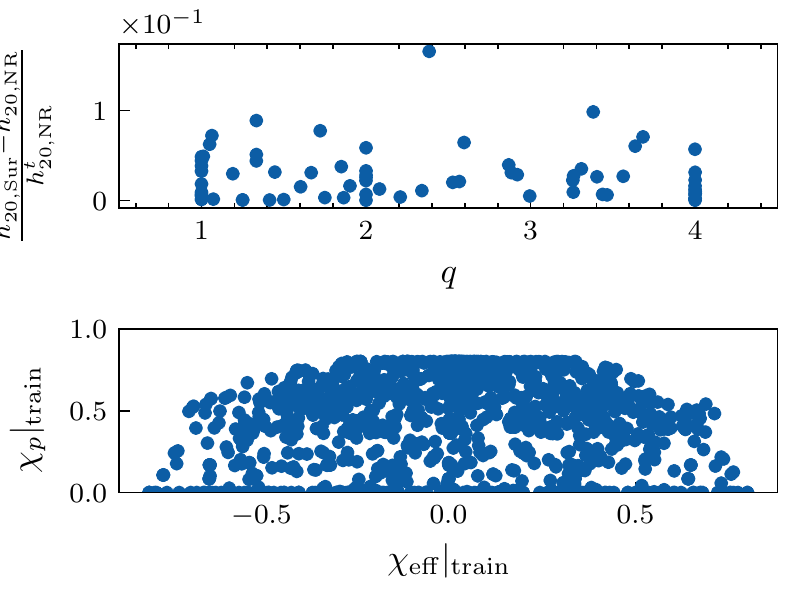}
    \caption{Top: The relativity difference between $h^t_{\rm 20Sur}$ and $h^t_{\rm 20NR}$ with respect to mass ratio $q$.  Bottom: The effective precession spin with respect to $\chi_{\rm eff}$.}\label{fig2BE}
\end{figure}

The effect of GW memory of BBH merger on the interferometory detector can be well described by \cite{Fav09b}
\begin{align}
h^{\rm mem}&=\frac{M}{D}\Re[(F^+(\theta,\phi,\psi)+iF^\times(\theta,\phi,\psi))\times\nonumber\\
&\sum_{l=2}^{\infty}\sum_{m=-l}^{l}h^t_{lm}Y_{-2lm}(\iota,\beta)]\nonumber\\
&\approx \frac{M}{D}F^+(\theta,\phi,\psi)h^t_{20}Y_{-220}(\iota),\label{eq2}\\
F^+(\theta,\phi,\psi)&\equiv-\frac{1}{2}(1+\cos^2\theta)\cos2\phi\cos2\psi\nonumber\\
&\,\,\,\,-\cos\theta\sin2\phi\sin2\psi,\\
F^\times(\theta,\phi,\psi)&\equiv+\frac{1}{2}(1+\cos^2\theta)\cos2\phi\sin2\psi\nonumber\\
&\,\,\,\,-\cos\theta\sin2\phi\cos2\psi,
\end{align}
where $\iota$ is the inclination angle of the BBH orbit plane respect to the observation direction, $\beta$ is the longitude angle describing the observation direction in the source frame, $(\theta,\phi)$ is the angular position of the BBH and $\psi$ is the polarization angle of the GW.
\begin{figure}
\begin{tabular}{c}
\includegraphics[width=0.5\textwidth]{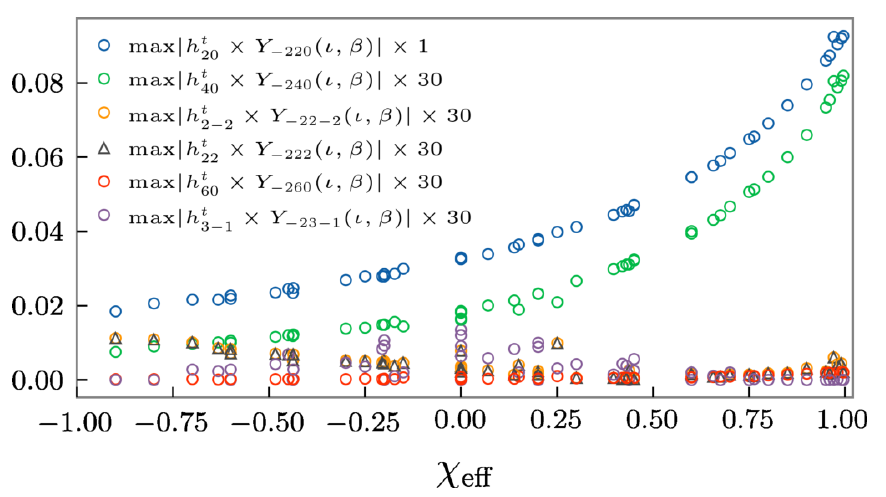}
\end{tabular}
\caption{Comparison of different memory modes contribution. The factor $Y_{-2lm}(\iota,\beta)$ corresponds to the effect of spin weighted spherical harmonic function. The maximal is taken respect to $\iota$ and $\beta$. This plot indicates that $(2,0)$ mode is greater than the next strongest mode $(4,0)$ more than 30 times.}\label{fig3}
\end{figure}

Because of the approximation of GW memory by just $(2,0)$ mode in (\ref{eq2}), the parameter $\beta$ can be ignored. Regarding to binary black hole coalescence systems the (2,0) mode overwhelmly dominates the GW memory.
The 6 leading contribution modes are compared in Fig.~\ref{fig3}. We can see that $(2,0)$ mode is stronger than the next strongest memory mode $(4,0)$ more than 30 times. Consequently the approximation Eq.~(\ref{eq2}) is safely satisfied. For each BBH system, $h^t_{20}$ is determined completely by the BBH intrinsic parameters $(q,\vec{\chi}_1,\vec{\chi}_2)$, where $q\geq1$ is the mass ratio. We have calculated the corresponding memory according to the Eq.~(\ref{meq4}) for 1370 simulations of generic, fully precessing BBHs with mass ratios $1\leq q\leq4$ and spin magnitudes $|\vec{\chi}_1|,|\vec{\chi}_2|<0.8$. The resulted memory data $h^t_{\rm 20NR}$ are available online\footnote{\url{https://github.com/Zhi-ChaoZhao/NRSurMemory\_7qd4/blob}\\
\url{/main/Data\_of\_Paper/Training\_Data.csv}}.

Based on the above 1370 GW memory results for generic fully precessing BBHs, we have constructed a surrogate model to describe the relation between the BBH intrinsic parameters $(q,\vec{\chi}_1,\vec{\chi}_2)$ and $h^t_{20}$. Our construction procedure closely follows \cite{PhysRevLett.122.011101,PhysRevResearch.1.033015}. Due to the precession, BH spins $\vec{\chi}_1,\vec{\chi}_2$ will change with time. We take the spin at time $t=-100M$ respect to the merger time as the initial parameters which is the same as \cite{PhysRevLett.122.011101,PhysRevResearch.1.033015}. We randomly choose 1301 samples among the above mentioned 1370 simulations to train and obtain a surrogate model \NRSurMemory. The rest 69 samples are used to check the accuracy of our model \NRSurMemory. The difference between $h^t_{\rm 20NR}$ and $h^t_{\rm 20Sur}$ for these 69 samples are plotted in the top panel of Fig.~\ref{fig2}. At the same time we plot the estimated error by our Gaussian process regression type model \NRSurMemory in the bottom panel of Fig.~\ref{fig2}. We can see the estimated error by the model \NRSurMemory is consistent to the difference between the model prediction and the numerical relativity result. This feature indicates the reliability of the \NRSurMemory model. The related python code and the model involved data of \NRSurMemory are also available online\footnote{\url{https://github.com/Zhi-ChaoZhao/NRSurMemory\_7qd4}}.

We also checked the dependence of errors in Fig.~\ref{fig2} with the mass ratio $q$ and the effective spin $\chi_{\rm eff}$, shown in Fig.~\ref{fig2A}.
We notice the errors seems to decrease with $q$, which is because the memory is larger when $q$ is smaller. We plot the relative error in the top panel of Fig.~\ref{fig2BE}, which shows the relative error is independent of $q$.
While the bottom panel of Fig.~\ref{fig2A} tells us that the error seems to be larger when $|\chi_{\rm eff}|$. That might comes from the non-uniform distribution of our training set.
We plot the distribution of effective precession spin with respect to $\chi_{\rm eff}$ in the bottom panel of Fig.~\ref{fig2BE}. Obviously, when $\chi_{\rm eff}$ near zero, the larger effective precession spin becomes more, whose memory may be harder to model.

\begin{figure}
\begin{tabular}{c}
\includegraphics[width=0.5\textwidth]{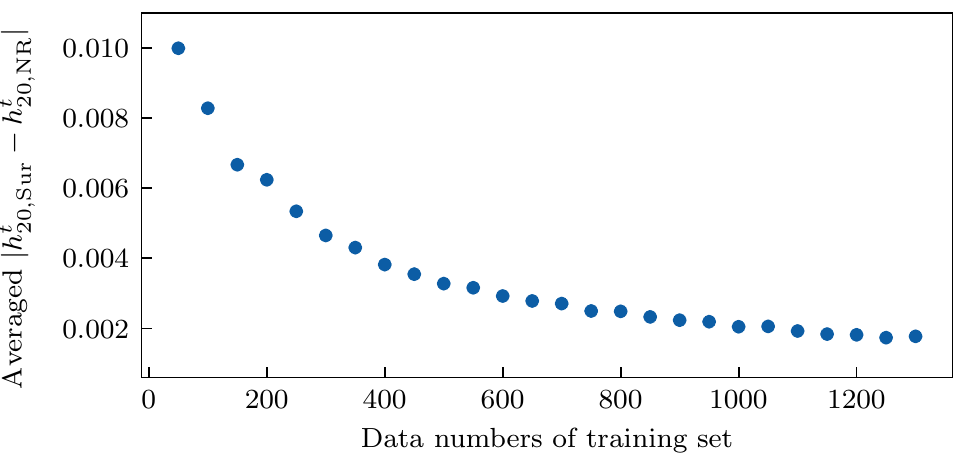}
\end{tabular}
\caption{The averaged accuracy of the surrogate model respect to the number of training data. We have used 50 runs and taken the corresponding average.}\label{fig4}
\end{figure}

Our surrogate model for memory is based on Gaussian process regression. Here we check the effect of the number of training data on the accuracy of the resulted surrogate model. In all we have 1370 numerical relativity results for GW memory based on SXS simulations. Denote the number of the training data $N<1370$. We randomly take $N$ samples from the 1370 results to train the surrogate model and use the rest $1370-N$ samples as test set. We repeat this process 50 times and average the resulted error. Here 50 is arbitrarily chosen and it does not affect the test result. We plot such averaged error respect to $N$ in Fig.~\ref{fig4}. We found that when
the number is larger than 1200, the results will not change any more. This is also the reason we chose 1301 samples for training above.

\section{GW memory estimation of the BBH events in GWTC-2}

\begin{table*}
\centering
\caption{The Kullback-Leibler (KL) divergence between the prior and posterior distribution for the GW memory strain $h^{\rm mem}$ affected on each detector of the 48 BBH events recorded in GWTC-2.}\label{table1}
\begin{tabular}{l|c|c|c||l|c|c|c}
\hline
           Event &  $D_{\rm KL,H1}$ &  $D_{\rm KL,L1}$ &  $D_{\rm KL,V1}$ & Event &  $D_{\rm KL,H1}$ &  $D_{\rm KL,L1}$ &  $D_{\rm KL,V1}$ \\
\hline
        GW150914 &            2.394 &            1.748 &     $\backslash$ &    GW190521\_074359 &            0.410 &            0.761 &     $\backslash$ \\
        GW151012 &            0.532 &            0.529 &     $\backslash$ &    GW190527\_092055 &            0.213 &            0.239 &     $\backslash$ \\
        GW151226 &            3.441 &            3.662 &     $\backslash$ &    GW190602\_175927 &            0.135 &            0.185 &            0.057 \\
        GW170104 &            1.307 &            1.138 &     $\backslash$ &    GW190620\_030421 &     $\backslash$ &            0.361 &            0.047 \\
        GW170608 &            5.060 &            2.278 &     $\backslash$ &    GW190630\_185205 &     $\backslash$ &            1.180 &            0.809 \\
        GW170729 &            0.272 &            0.218 &            0.017 &    GW190701\_203306 &            0.112 &            0.127 &            0.603 \\
        GW170809 &            0.488 &            0.745 &            0.041 &    GW190706\_222641 &            0.303 &            0.149 &            0.021 \\
        GW170814 &            1.487 &            1.555 &            1.315 &    GW190707\_093326 &            0.363 &            0.495 &     $\backslash$ \\
        GW170818 &            0.515 &            0.918 &            1.378 &    GW190708\_232457 &     $\backslash$ &            0.723 &            0.025 \\
        GW170823 &            0.205 &            0.227 &     $\backslash$ &    GW190719\_215514 &            0.137 &            0.152 &     $\backslash$ \\
 GW190408\_181802 &            0.104 &            0.065 &            0.459 &    GW190720\_000836 &            4.636 &            3.508 &            3.310 \\
        GW190412 &            0.460 &            0.869 &            1.214 &    GW190727\_060333 &            0.342 &            0.211 &            0.018 \\
 GW190413\_052954 &            0.132 &            0.132 &            0.090 &    GW190728\_064510 &            3.946 &            4.622 &            0.546 \\
 GW190413\_134308 &            0.074 &            0.049 &            0.144 &    GW190731\_140936 &            0.077 &            0.037 &     $\backslash$ \\
 GW190421\_213856 &            0.054 &            0.024 &     $\backslash$ &    GW190803\_022701 &            0.061 &            0.058 &            0.017 \\
 GW190424\_180648 &     $\backslash$ &            1.795 &     $\backslash$ &           GW190814 &            1.782 &            2.195 &            1.422 \\
 GW190426\_152155 &            0.970 &            0.383 &            0.186 &    GW190828\_063405 &            0.767 &            0.394 &            0.081 \\
 GW190503\_185404 &            0.171 &            0.049 &            0.079 &    GW190828\_065509 &            0.109 &            0.069 &            0.256 \\
 GW190512\_180714 &            0.375 &            0.707 &            0.076 &    GW190909\_114149 &            0.208 &            0.160 &     $\backslash$ \\
 GW190513\_205428 &            0.140 &            0.119 &            0.216 &    GW190910\_112807 &     $\backslash$ &            2.641 &            0.233 \\
 GW190514\_065416 &            0.064 &            0.053 &     $\backslash$ &    GW190915\_235702 &            0.282 &            0.088 &            0.063 \\
 GW190517\_055101 &            0.604 &            0.668 &            0.609 &    GW190924\_021846 &            0.896 &            1.665 &            0.097 \\
 GW190519\_153544 &            2.920 &            2.736 &            0.309 &    GW190929\_012149 &            0.471 &            0.492 &            0.065 \\
        GW190521 &            0.114 &            0.113 &            0.031 &    GW190930\_133541 &            1.459 &            1.861 &     $\backslash$ \\
\hline
\end{tabular}
\end{table*}

\begin{figure*}
\includegraphics[width=\textwidth]{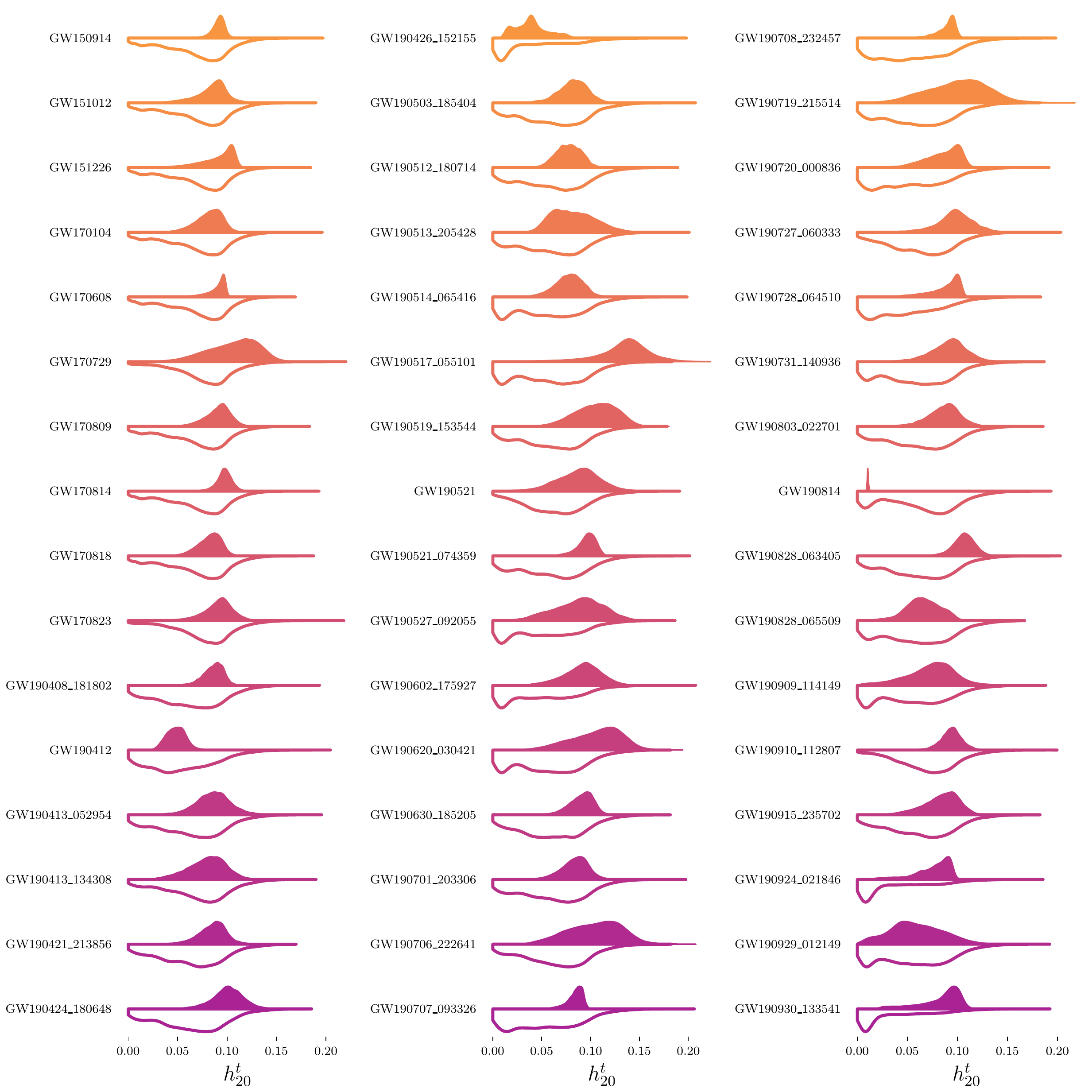}
\caption{The posterior probability and the prior probability of $(2,0)$ mode GW memory for the 48 BBH events recorded in GWTC-2.}\label{fig5}
\end{figure*}

\begin{figure*}
\begin{tabular}{c}
\includegraphics[width=0.62\textwidth]{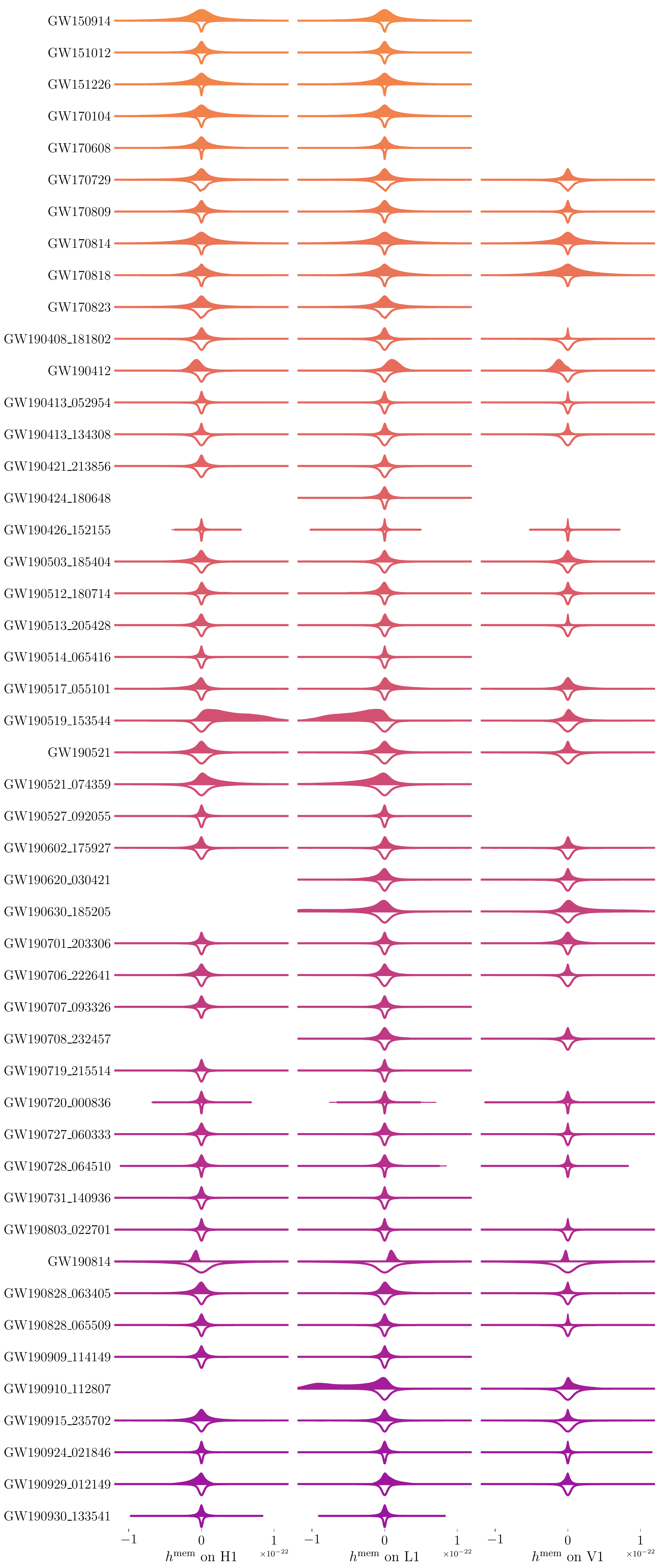}
\end{tabular}
\caption{The posterior probability and the prior probability of memory for the 48 BBH events recorded in GWTC-2 for the three detectors. For some events, some detectors were not working properly where the plot is absent.}\label{fig6}
\end{figure*}

\begin{figure}
\includegraphics[width=0.5\textwidth]{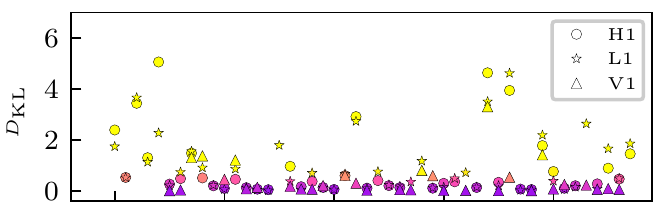}
\caption{KL divergence of the GW memory estimation for the 48 BBH events in GWTC-2. The horizontal axis corresponds to the 48 BBH events. For most events there are three estimation results corresponding to the three detectors, LIGO Hanford (H1), LIGO Livingston (L1) and VIRGO (V1).}\label{fig7}
\end{figure}

\begin{figure*}
\begin{tabular}{c}
\includegraphics[width=\textwidth]{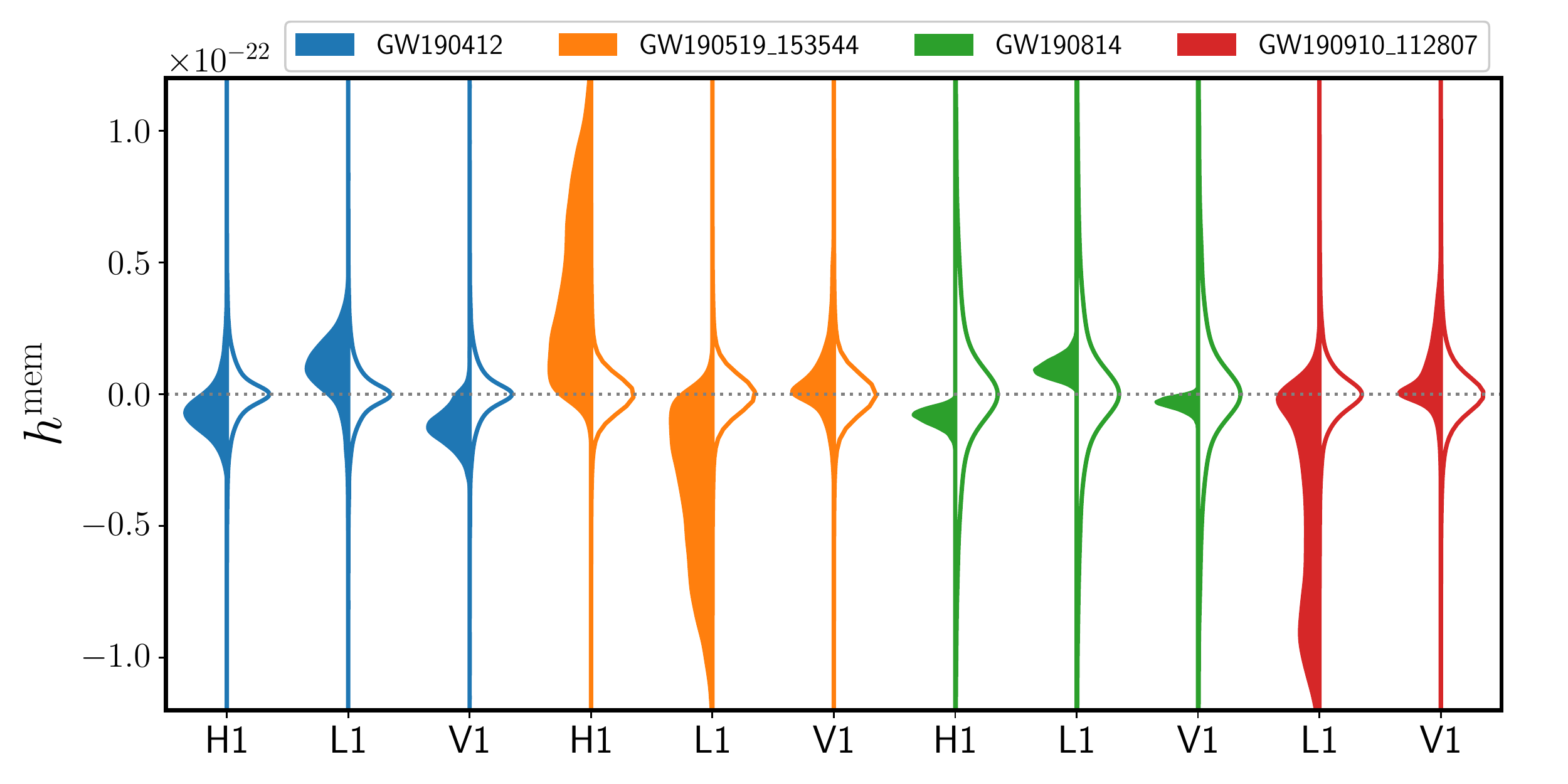}
\end{tabular}
\caption{Violin plot for the prior (right) and posterior (left) distribution of the GW memory $h^{\rm mem}$ for four BBH events in GWTC-2. The memory of these four BBH events has been estimated most accurately among the 48 BBH events of GWTC-2. The memory is respect to specific detector. So each plot responds to a specific detector. During the time of the event GW190910\_112807, H1 detector did not work well, so the corresponding plot is absent.}\label{fig8}
\end{figure*}
Given a distribution probability of parameters $(M,q,\vec{\chi}_1,\vec{\chi}_2,D,\iota,\theta,\phi,\psi)$ for a BBH system, our GW memory model (\ref{eq2}) can result in an estimation of GW memory with a corresponding probability. For each detected BBH system by LIGO and VIRGO, the parameters $(M,q,\vec{\chi}_1,\vec{\chi}_2,D,\iota,\theta,\phi,\psi)$ can be estimated with a posterior probability based on a given prior probability \cite{PhysRevLett.124.101104}. So we can accordingly estimate the GW memory for each BBH event in GWTC-2. At the same time we can also present the corresponding prior and posterior probability for the GW memory. Different to our work, the authors in \cite{PhysRevD.103.044012} estimated the intrinsic factor $h^t_{20}$ only through the intrinsic parameters $(q,\vec{\chi}_1,\vec{\chi}_2)$ for O1/O2 BBH events.

The 38 BBH events during O3a haven been announced in GWTC-2 \cite{2020arXiv201014527A} by LIGO Scientific Collaboration (LSC). The 10 BBH events for O1/O2 reported in the GWTC-1 \cite{PhysRevX.9.031040} do not include the full information of black hole spin. We use the analysis results by the Bilby group \cite{10.1093.mnras.staa2850} where the full information is available. We infer the GW memory based on these two analysis results.

We firstly plot the posterior probability and the prior probability of $(2,0)$ mode GW memory $h^t_{20}$ for the 48 BBH events recorded in GWTC-2 in Fig.~\ref{fig5}. Note that our $h^t_{20}$ is different to $\Delta\mathfrak{h}_{20}$ of \cite{PhysRevD.103.044012}, $h^t_{20}=\frac{D}{M}\Delta\mathfrak{h}_{20}$. For all events the posterior distribution of $h^t_{20}$ is clearly different to that of the prior distribution. This means we have already well estimated $(2,0)$ mode GW memory for the 48 BBH events. But this quantity can not be related to any detection directly. That is why we consider inferring the GW memory strain $h^{\rm mem}$ affected on each detector in the current work.

We plot the prior distribution and the posterior distribution for $h^{\rm mem}$ of the 48 BBH events respectively in Fig.~\ref{fig6}. The GW memory means the permanent change of the gravitational wave strain affected on each detector. So there are three plots corresponding to each BBH event. If the specific detector did not work properly when a BBH event happen, the corresponding plot is absent.

In order to quantify how much information our GW memory estimation has got from the gravitational wave detection, we investigate the Kullback-Leibler (KL) divergence between the prior and posterior distribution. We plot the resulted KL divergences for the memory estimation respect to the 48 BBH events in Fig.~\ref{fig7} and list the corresponding KL divergence in Table.~\ref{table1}. There are 16 events admitting KL divergence bigger than 1. These big KL divergences indicate that good information has been obtained by our GW memory estimation.

Among the above mentioned 16 events with good memory estimation, we find that GW190412, GW190519\_153544, GW190814 and GW190910\_112807 admit clear nonvanishing mean values for the posterior distributions of the GW memory strain $h^{\rm mem}$ affected on each detector. All prior distributions are approximated Gaussian distribution with vanishing mean value. If the detection data gives little information to the memory, approximated Gaussian posterior distribution with vanishing mean value will be resulted. And correspondingly small KL divergence will be got. In contrast, if the detection data introduce significant information to the memory, the posterior distribution will admit nonvanishing mean value. GW190412, GW190519\_153544, GW190814 and GW190910\_112807 fall in this category. We show the violin plot for the prior and posterior distribution of the GW memory strain in Fig.~\ref{fig8}. Due to the configurations of H1 and L1, we know the GW memory strains on H1 and L1 admit different signs. But Fig.~\ref{fig8} definitely tells us the signs of the GW memory affected on H1 and L1 for the first time.

\section{The effect of waveform model on the estimation of gravitational wave memory}
We surely know that the gravitational waveform model may affect the parameters estimation. People call such effect the “systematic bias of the waveform template \cite{PhysRevD.94.124030,PhysRevResearch.2.023151}. The most typical example is GW190521. Significant difference shows up among SEOBNR waveform families, IMRPhenom waveform families and NRsurrogate waveform model. These three waveform families are the most advanced waveform templates available to gravitational wave data analysis. The reported analysis results by LSC have already counted the waveform accuracy issue. In addition, LSC has also considered the waveform models combination and prior distribution of related parameters based on detail astrophysical issues. Since the parameters estimation done by LSC group is extremely delicate, the posterior samples given by LSC are the most reasonable start point for the application of our technique to the LIGO GW events. This is the guide idea for the GW memory estimation done in the above section.

But it is still interesting to ask how the waveform model and the prior distribution of related parameters affect the GW memory estimation. We do such investigation here.

\subsection{The impact of waveform model systematics}
In the GWTC-2 paper \cite{ligo2020gwtc}, LSC group has released the posterior distribution of GW source parameters. The result of
GW190412 used the combination of IMRPhenomPv3HM and SEOBNRv4PHM; the result of GW190519\_153544 used the
combination of NRSur7dq4 and SEOBNRv4PHM; the result of GW190814 used the combination of IMRPhenomPv3HM and
SEOBNRv4PHM; and the result of GW190910\_112807 used SEOBNRv4PHM.

\begin{figure*}
    \centering
    \includegraphics[width=\textwidth]{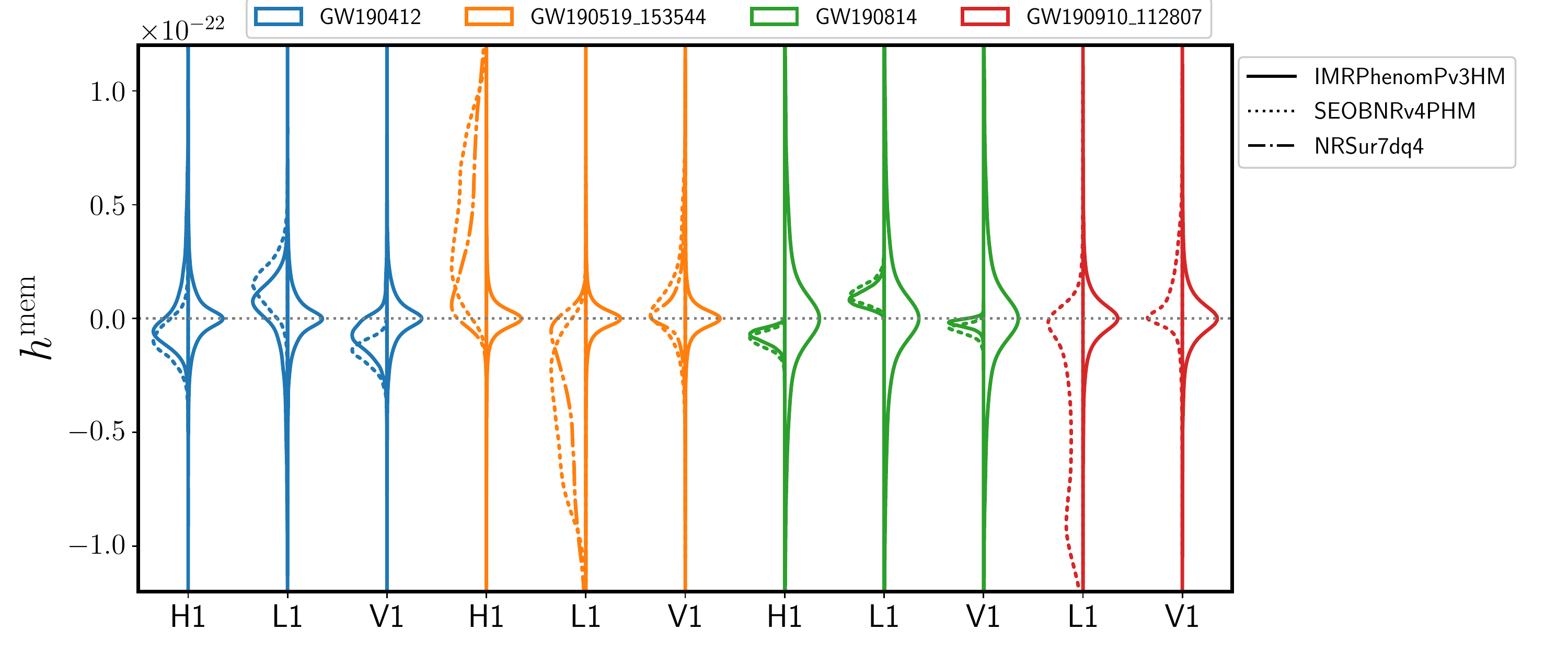}
    \caption{GW memory estimations based on different waveform models. For each violin plot, the left panel corresponds to
    the posterior distribution and the right panel corresponds to the prior distribution. The three waveform models are IMRPhenomPv3HM,
    SEOBNRv4PHM, and NRSur7dq4 respectively which have been listed in the legend.}
    \label{fig9}
\end{figure*}

In order to explore the effect of waveform models on our GW memory estimation, we use the individual posterior of the above mentioned waveform models to estimate GW memory
instead of using the combined posterior. The results are plotted in Fig.~\ref{fig9}. In this
figure, we can see that different waveform models may affect the GW memory estimation. Just as the effect of waveform models on GW source parameters estimation \cite{2020arXiv201014527A}, the result shown in Fig.~\ref{fig9} is consistent to our expectation. But we would like to emphasize that the results reported in the above section are robust to waveform model choices. The interesting features of the GW memory of the four events do not change. The corresponding GW memory of GW190814 is always about $-1\times10^{-23}$ and $1\times10^{-23}$ for Hanford detector and Livingston detector respectively which is independent of waveform models.

\subsection{The effect of prior assumptions}
According to Bayesian theorem, prior distribution may affect the posterior distribution. Here we check how the prior distribution assumptions would
affect the GW memory estimation.

\begin{figure*}
    \centering
    \includegraphics[width=\textwidth]{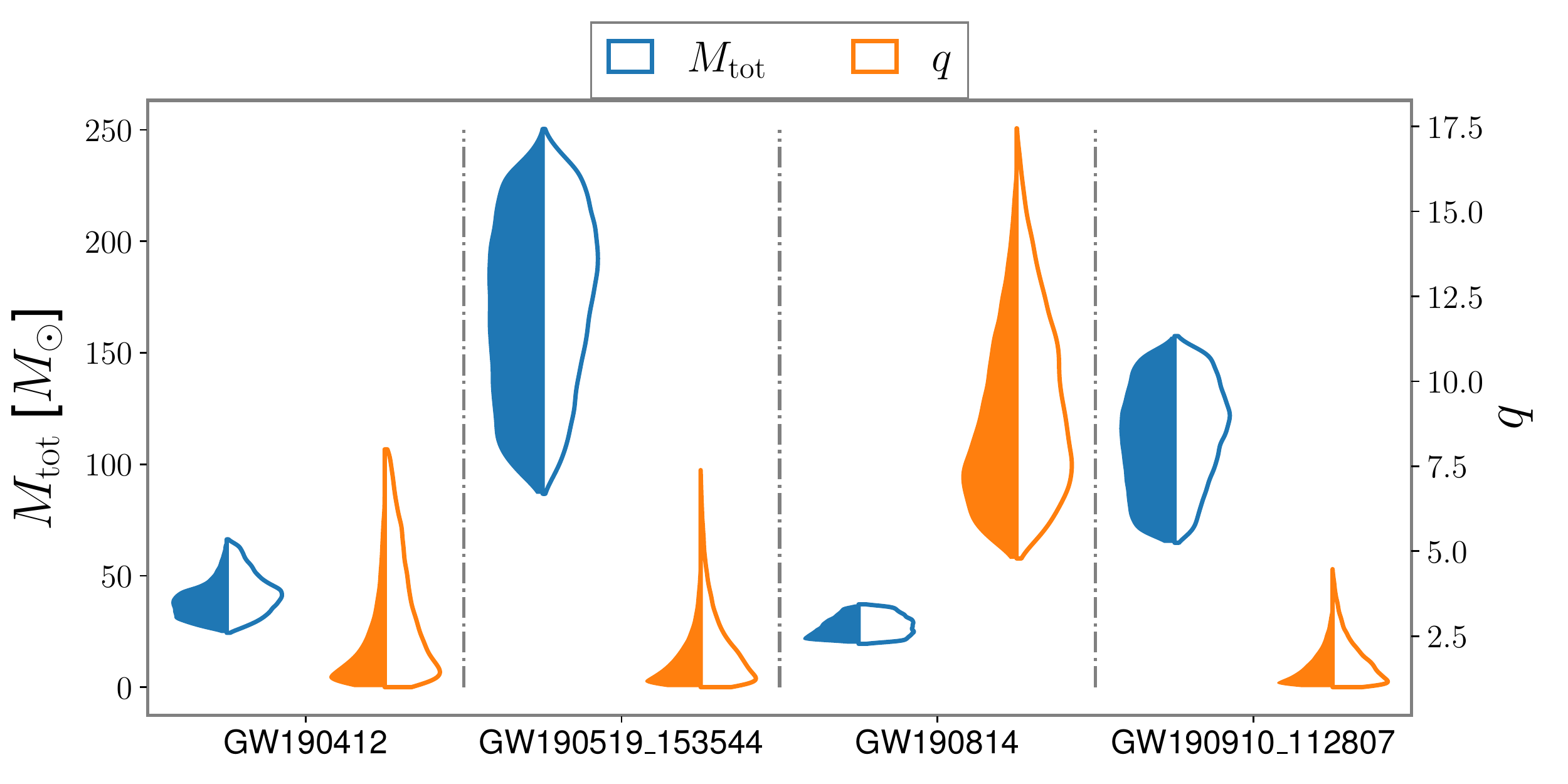}
    \caption{Prior distribution of the total mass $M_{\rm tot}$ and the mass ratio $q$ of the binary. The left panel of each violin plot corresponding to Prior A, and the right panel
    of each violin plot corresponding to Prior B. The blue violin plots are for $M_{\rm tot}$, and the orange plots are for $q$.}
    \label{fig10}
\end{figure*}

We use IMRPhenomXPHM \cite{garcia2020multimode} as the waveform model to check the dependence of GW memory estimation on the prior distribution. We use two different priors for the comparison.
For both prior distributions, we take uniform distribution for the BH's spin magnitudes and isotropic distribution for the BH's spin orientations, binary's sky location and the orbital orientation. The prior distribution of the luminosity distance corresponds to a uniform merger rate in the co-moving frame of the source. The difference of the two priors is about the masses of the two components.
With prior A, we assume that the chirp mass is uniformly distributed. With prior B, we assume that the component masses of the binary is uniformly distributed. We plot the resulted distribution of the total mass $M_{\rm tot}$ and mass ratio $q$ in Fig.~\ref{fig10} corresponding respectively to the two priors. From Fig.~\ref{fig10} we can see the two prior distributions respect to the total mass $M_{\rm tot}$ and mass ratio $q$ are significantly different to each other.

\begin{figure*}
    \centering
    \includegraphics[width=\textwidth]{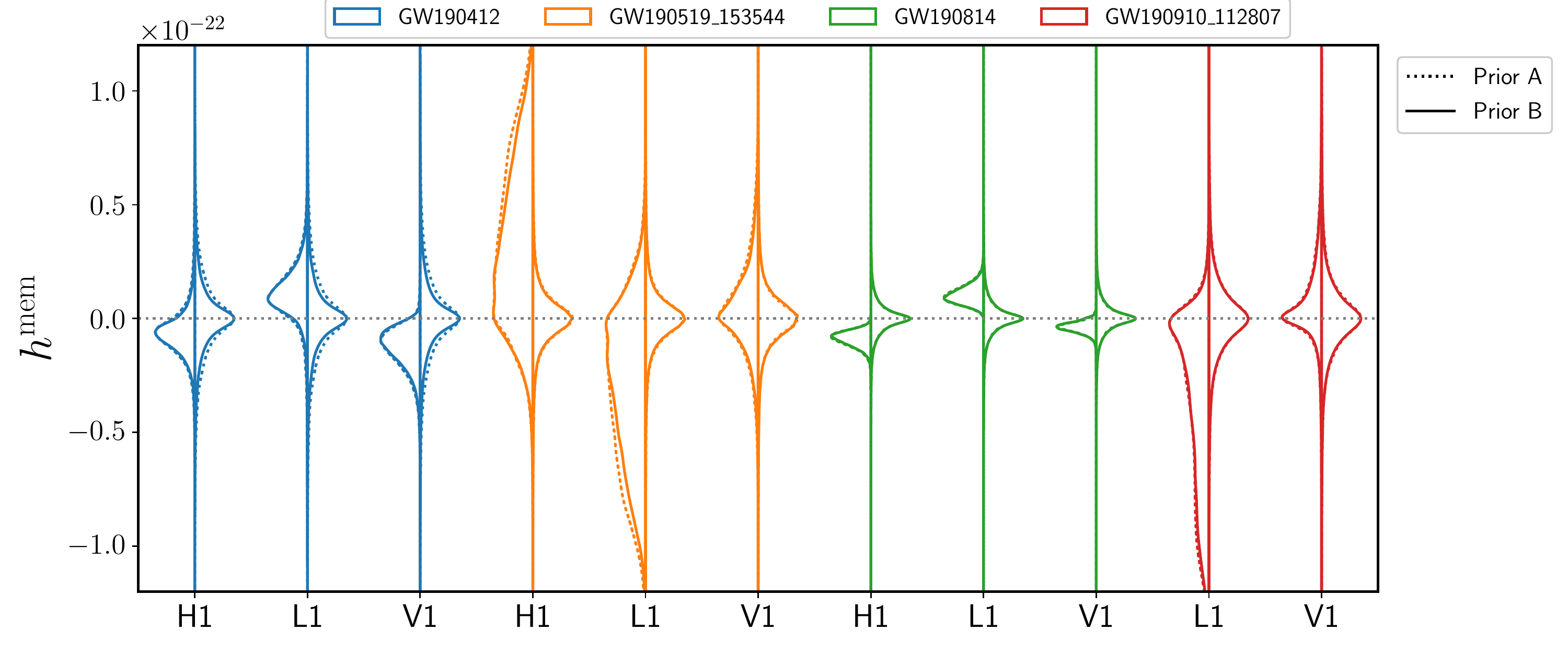}
    \caption{GW memory estimations based on different prior distributions shown in Fig.~\ref{fig10}. For each violin plot, the left panel corresponds to
    the posterior distribution and the right panel corresponds to the prior distribution. The corresponding prior distributions have been listed in the legend.}
    \label{fig11}
\end{figure*}

We plot the resulted GW memory estimation based on the two different prior distributions in Fig.~\ref{fig11}. From this figure we can see the two GW memory estimation results are roughly the same even the prior distributions are significantly different as shown in Fig.~\ref{fig10}. Such independent behavior of GW memory estimation on the prior distribution is consistent with the high KL divergence result got in the above section.

Based on the above analysis, we conclude that the estimated GW memory shown in Fig.~\ref{fig8} is robust to the waveform models and the prior assumptions.

\section{Summary and discussion}
We have implemented a completely different GW memory measurement scheme (more precisely, estimation scheme\cite{PhysRevLett.122.011101}) compared to the existing works in the literature. In order to realize our measurement scheme, we have applied the Bondi-Metzner-Sachs method to accurately calculate the GW memory for BBH. Combining this method and the SXS numerical relativity simulation we construct a database to relate BBH initial parameters and the corresponding GW memory. Aided with this database we have constructed a Gaussian process regression type surrogate model \NRSurMemory for GW memory of BBH. With this powerful model, we have done an estimation of the GW memory for the 48 BBH events of GWTC-2. Different to the GW memory waveform models \cite{PhysRevD.98.064031,PhysRevD.103.043005}, our surrogate model \NRSurMemory describes the overall GW memory instead of the waveform.

Previous GW memory measurements are all based on waveform analysis. There is no hope to detect GW memory in the near future with such waveform analysis method \cite{PhysRevD.101.023011}. The measurement method used in the current work is completely different \cite{PhysRevLett.124.101104}. The key bases for the current method are the Bondi-Metzner-Sachs GW memory calculation technique and the powerful model \NRSurMemory.

Different to previous qualitative estimate on the strength of GW memory \cite{PhysRevLett.118.181103}, we present the first quantitative measurement of GW memory for the 48 BBH events in GWTC-2. Together with the median value, the posterior distribution of GW memory is also presented. According to the KL divergence between the prior distribution and the posterior distribution, we found 16 GW memory measurements are trustable. This feature is different to the behavior of kick velocity \cite{PhysRevLett.124.101104}. More interestingly we found 4 GW memory measurements definitely tell the signs of the memory on LIGO detectors. Among them we are sure with more than 99.979\% confidence that the memory strain of GW190814 is negative on Hanford detector while positive on Livingston detector. In the future, when other GW memory detection results are available \cite{PhysRevD.101.023011}, the comparison to our estimation can strongly constrain general relativity \cite{PhysRevD.94.104063}.

Our estimation technique and our estimation results presented in the current paper can guide
people to more suitably choose the GW events for memory detection with multiple events.
Aided with our estimation technique, the GW memory detection method
with multiple events \cite{PhysRevLett.118.181103} will become
easier to detect GW memory. In addition, the GW memory features founded in
our work can be used to strongly test general relativity together with the
future GW memory detection.

\section*{Acknowledgments}
We thank Dr. He Wang for helpful discussions. This work was supported by the NSFC (No.~11690023 and No.~11920101003). Z. Cao was supported by ``the Interdiscipline Research Funds of Beijing Normal University" and the Strategic Priority Research Program of the Chinese Academy of Sciences, grant No. XDB23040100. X. He was supported by NSF of Hunan province (2018JJ2073).
\bibliography{refs}

\begin{thebibliography}{48}
\expandafter\ifx\csname natexlab\endcsname\relax\def\natexlab#1{#1}\fi
\expandafter\ifx\csname bibnamefont\endcsname\relax
  \def\bibnamefont#1{#1}\fi
\expandafter\ifx\csname bibfnamefont\endcsname\relax
  \def\bibfnamefont#1{#1}\fi
\expandafter\ifx\csname citenamefont\endcsname\relax
  \def\citenamefont#1{#1}\fi
\expandafter\ifx\csname url\endcsname\relax
  \def\url#1{\texttt{#1}}\fi
\expandafter\ifx\csname urlprefix\endcsname\relax\def\urlprefix{URL }\fi
\providecommand{\bibinfo}[2]{#2}
\providecommand{\eprint}[2][]{\url{#2}}

\bibitem[{\citenamefont{Zel'dovich and Polnarev}(1974)}]{Zeldovich74}
\bibinfo{author}{\bibfnamefont{Y.~B.} \bibnamefont{Zel'dovich}}
  \bibnamefont{and} \bibinfo{author}{\bibfnamefont{A.~G.}
  \bibnamefont{Polnarev}}, \bibinfo{journal}{Soviet Astronomy}
  \textbf{\bibinfo{volume}{18}}, \bibinfo{pages}{17} (\bibinfo{year}{1974}).

\bibitem[{\citenamefont{Payne}(1983)}]{Pay83}
\bibinfo{author}{\bibfnamefont{P.~N.} \bibnamefont{Payne}},
  \bibinfo{journal}{Phys. Rev. D} \textbf{\bibinfo{volume}{28}},
  \bibinfo{pages}{1894} (\bibinfo{year}{1983}).

\bibitem[{\citenamefont{Braginsky and Grishchuk}(1985)}]{Braginsky:1986ia}
\bibinfo{author}{\bibfnamefont{V.~B.} \bibnamefont{Braginsky}}
  \bibnamefont{and} \bibinfo{author}{\bibfnamefont{L.~P.}
  \bibnamefont{Grishchuk}}, \bibinfo{journal}{Sov. Phys. JETP}
  \textbf{\bibinfo{volume}{62}}, \bibinfo{pages}{427} (\bibinfo{year}{1985}).

\bibitem[{\citenamefont{Braginsky and
  Thorne}(1987)}]{braginsky1987gravitational}
\bibinfo{author}{\bibfnamefont{V.~B.} \bibnamefont{Braginsky}}
  \bibnamefont{and} \bibinfo{author}{\bibfnamefont{K.~S.}
  \bibnamefont{Thorne}}, \bibinfo{journal}{Nature}
  \textbf{\bibinfo{volume}{327}}, \bibinfo{pages}{123} (\bibinfo{year}{1987}).

\bibitem[{\citenamefont{Christodoulou}(1991)}]{PhysRevLett.67.1486}
\bibinfo{author}{\bibfnamefont{D.}~\bibnamefont{Christodoulou}},
  \bibinfo{journal}{Phys. Rev. Lett.} \textbf{\bibinfo{volume}{67}},
  \bibinfo{pages}{1486} (\bibinfo{year}{1991}).

\bibitem[{\citenamefont{Frauendiener}(1992)}]{Fra92}
\bibinfo{author}{\bibfnamefont{J.}~\bibnamefont{Frauendiener}},
  \bibinfo{journal}{Classical and Quantum Gravity}
  \textbf{\bibinfo{volume}{9}}, \bibinfo{pages}{1639} (\bibinfo{year}{1992}).

\bibitem[{\citenamefont{Yang and Martynov}(2018)}]{PhysRevLett.121.071102}
\bibinfo{author}{\bibfnamefont{H.}~\bibnamefont{Yang}} \bibnamefont{and}
  \bibinfo{author}{\bibfnamefont{D.}~\bibnamefont{Martynov}},
  \bibinfo{journal}{Phys. Rev. Lett.} \textbf{\bibinfo{volume}{121}},
  \bibinfo{pages}{071102} (\bibinfo{year}{2018}).

\bibitem[{\citenamefont{H\"ubner et~al.}(2020)\citenamefont{H\"ubner, Talbot,
  Lasky, and Thrane}}]{PhysRevD.101.023011}
\bibinfo{author}{\bibfnamefont{M.}~\bibnamefont{H\"ubner}},
  \bibinfo{author}{\bibfnamefont{C.}~\bibnamefont{Talbot}},
  \bibinfo{author}{\bibfnamefont{P.~D.} \bibnamefont{Lasky}}, \bibnamefont{and}
  \bibinfo{author}{\bibfnamefont{E.}~\bibnamefont{Thrane}},
  \bibinfo{journal}{Phys. Rev. D} \textbf{\bibinfo{volume}{101}},
  \bibinfo{pages}{023011} (\bibinfo{year}{2020}).

\bibitem[{\citenamefont{Boersma et~al.}(2020)\citenamefont{Boersma, Nichols,
  and Schmidt}}]{PhysRevD.101.083026}
\bibinfo{author}{\bibfnamefont{O.~M.} \bibnamefont{Boersma}},
  \bibinfo{author}{\bibfnamefont{D.~A.} \bibnamefont{Nichols}},
  \bibnamefont{and} \bibinfo{author}{\bibfnamefont{P.}~\bibnamefont{Schmidt}},
  \bibinfo{journal}{Phys. Rev. D} \textbf{\bibinfo{volume}{101}},
  \bibinfo{pages}{083026} (\bibinfo{year}{2020}).

\bibitem[{\citenamefont{Du and Nishizawa}(2016)}]{PhysRevD.94.104063}
\bibinfo{author}{\bibfnamefont{S.~M.} \bibnamefont{Du}} \bibnamefont{and}
  \bibinfo{author}{\bibfnamefont{A.}~\bibnamefont{Nishizawa}},
  \bibinfo{journal}{Phys. Rev. D} \textbf{\bibinfo{volume}{94}},
  \bibinfo{pages}{104063} (\bibinfo{year}{2016}).

\bibitem[{\citenamefont{Hollands et~al.}(2017)\citenamefont{Hollands,
  Ishibashi, and Wald}}]{Hollands_2017}
\bibinfo{author}{\bibfnamefont{S.}~\bibnamefont{Hollands}},
  \bibinfo{author}{\bibfnamefont{A.}~\bibnamefont{Ishibashi}},
  \bibnamefont{and} \bibinfo{author}{\bibfnamefont{R.~M.} \bibnamefont{Wald}},
  \bibinfo{journal}{Classical and Quantum Gravity}
  \textbf{\bibinfo{volume}{34}}, \bibinfo{pages}{155005}
  (\bibinfo{year}{2017}).

\bibitem[{\citenamefont{Seto}(2009)}]{Set09}
\bibinfo{author}{\bibfnamefont{N.}~\bibnamefont{Seto}},
  \bibinfo{journal}{Monthly Notices of the Royal Astronomical Society: Letters}
  \textbf{\bibinfo{volume}{400}}, \bibinfo{pages}{L38} (\bibinfo{year}{2009}).

\bibitem[{\citenamefont{Van~Haasteren and Levin}(2010)}]{VanLev10}
\bibinfo{author}{\bibfnamefont{R.}~\bibnamefont{Van~Haasteren}}
  \bibnamefont{and} \bibinfo{author}{\bibfnamefont{Y.}~\bibnamefont{Levin}},
  \bibinfo{journal}{Monthly Notices of the Royal Astronomical Society}
  \textbf{\bibinfo{volume}{401}}, \bibinfo{pages}{2372} (\bibinfo{year}{2010}).

\bibitem[{\citenamefont{Pshirkov et~al.}(2010)\citenamefont{Pshirkov, Baskaran,
  and Postnov}}]{PshBasPos10}
\bibinfo{author}{\bibfnamefont{M.}~\bibnamefont{Pshirkov}},
  \bibinfo{author}{\bibfnamefont{D.}~\bibnamefont{Baskaran}}, \bibnamefont{and}
  \bibinfo{author}{\bibfnamefont{K.}~\bibnamefont{Postnov}},
  \bibinfo{journal}{Monthly Notices of the Royal Astronomical Society}
  \textbf{\bibinfo{volume}{402}}, \bibinfo{pages}{417} (\bibinfo{year}{2010}).

\bibitem[{\citenamefont{Cordes and Jenet}(2012)}]{CorJen12}
\bibinfo{author}{\bibfnamefont{J.}~\bibnamefont{Cordes}} \bibnamefont{and}
  \bibinfo{author}{\bibfnamefont{F.}~\bibnamefont{Jenet}},
  \bibinfo{journal}{The Astrophysical Journal} \textbf{\bibinfo{volume}{752}},
  \bibinfo{pages}{54} (\bibinfo{year}{2012}).

\bibitem[{\citenamefont{Madison et~al.}(2014)\citenamefont{Madison, Cordes, and
  Chatterjee}}]{MadCorCha14}
\bibinfo{author}{\bibfnamefont{D.}~\bibnamefont{Madison}},
  \bibinfo{author}{\bibfnamefont{J.}~\bibnamefont{Cordes}}, \bibnamefont{and}
  \bibinfo{author}{\bibfnamefont{S.}~\bibnamefont{Chatterjee}},
  \bibinfo{journal}{The Astrophysical Journal} \textbf{\bibinfo{volume}{788}},
  \bibinfo{pages}{141} (\bibinfo{year}{2014}).

\bibitem[{\citenamefont{Arzoumanian et~al.}(2015)\citenamefont{Arzoumanian,
  Brazier, Burke-Spolaor, Chamberlin, Chatterjee, Christy, Cordes, Cornish,
  Demorest, Deng et~al.}}]{Arzoumanian_2015}
\bibinfo{author}{\bibfnamefont{Z.}~\bibnamefont{Arzoumanian}},
  \bibinfo{author}{\bibfnamefont{A.}~\bibnamefont{Brazier}},
  \bibinfo{author}{\bibfnamefont{S.}~\bibnamefont{Burke-Spolaor}},
  \bibinfo{author}{\bibfnamefont{S.~J.} \bibnamefont{Chamberlin}},
  \bibinfo{author}{\bibfnamefont{S.}~\bibnamefont{Chatterjee}},
  \bibinfo{author}{\bibfnamefont{B.}~\bibnamefont{Christy}},
  \bibinfo{author}{\bibfnamefont{J.~M.} \bibnamefont{Cordes}},
  \bibinfo{author}{\bibfnamefont{N.~J.} \bibnamefont{Cornish}},
  \bibinfo{author}{\bibfnamefont{P.~B.} \bibnamefont{Demorest}},
  \bibinfo{author}{\bibfnamefont{X.}~\bibnamefont{Deng}}, \bibnamefont{et~al.},
  \bibinfo{journal}{The Astrophysical Journal} \textbf{\bibinfo{volume}{810}},
  \bibinfo{pages}{150} (\bibinfo{year}{2015}).

\bibitem[{\citenamefont{Lasky et~al.}(2016)\citenamefont{Lasky, Thrane, Levin,
  Blackman, and Chen}}]{PhysRevLett.117.061102}
\bibinfo{author}{\bibfnamefont{P.~D.} \bibnamefont{Lasky}},
  \bibinfo{author}{\bibfnamefont{E.}~\bibnamefont{Thrane}},
  \bibinfo{author}{\bibfnamefont{Y.}~\bibnamefont{Levin}},
  \bibinfo{author}{\bibfnamefont{J.}~\bibnamefont{Blackman}}, \bibnamefont{and}
  \bibinfo{author}{\bibfnamefont{Y.}~\bibnamefont{Chen}},
  \bibinfo{journal}{Phys. Rev. Lett.} \textbf{\bibinfo{volume}{117}},
  \bibinfo{pages}{061102} (\bibinfo{year}{2016}).

\bibitem[{\citenamefont{McNeill et~al.}(2017)\citenamefont{McNeill, Thrane, and
  Lasky}}]{PhysRevLett.118.181103}
\bibinfo{author}{\bibfnamefont{L.~O.} \bibnamefont{McNeill}},
  \bibinfo{author}{\bibfnamefont{E.}~\bibnamefont{Thrane}}, \bibnamefont{and}
  \bibinfo{author}{\bibfnamefont{P.~D.} \bibnamefont{Lasky}},
  \bibinfo{journal}{Phys. Rev. Lett.} \textbf{\bibinfo{volume}{118}},
  \bibinfo{pages}{181103} (\bibinfo{year}{2017}).

\bibitem[{\citenamefont{Divakarla et~al.}(2020)\citenamefont{Divakarla, Thrane,
  Lasky, and Whiting}}]{PhysRevD.102.023010}
\bibinfo{author}{\bibfnamefont{A.~K.} \bibnamefont{Divakarla}},
  \bibinfo{author}{\bibfnamefont{E.}~\bibnamefont{Thrane}},
  \bibinfo{author}{\bibfnamefont{P.~D.} \bibnamefont{Lasky}}, \bibnamefont{and}
  \bibinfo{author}{\bibfnamefont{B.~F.} \bibnamefont{Whiting}},
  \bibinfo{journal}{Phys. Rev. D} \textbf{\bibinfo{volume}{102}},
  \bibinfo{pages}{023010} (\bibinfo{year}{2020}).

\bibitem[{\citenamefont{Caltech-Cornell-CITA}()}]{SXSBBH}
\bibinfo{author}{\bibnamefont{Caltech-Cornell-CITA}},
  \emph{\bibinfo{title}{binary black hole simulation results}},
  \bibinfo{howpublished}{\url{http://www.black-holes.org/waveforms}}.

\bibitem[{\citenamefont{Khera et~al.}(2021)\citenamefont{Khera, Krishnan,
  Ashtekar, and De~Lorenzo}}]{PhysRevD.103.044012}
\bibinfo{author}{\bibfnamefont{N.}~\bibnamefont{Khera}},
  \bibinfo{author}{\bibfnamefont{B.}~\bibnamefont{Krishnan}},
  \bibinfo{author}{\bibfnamefont{A.}~\bibnamefont{Ashtekar}}, \bibnamefont{and}
  \bibinfo{author}{\bibfnamefont{T.}~\bibnamefont{De~Lorenzo}},
  \bibinfo{journal}{Phys. Rev. D} \textbf{\bibinfo{volume}{103}},
  \bibinfo{pages}{044012} (\bibinfo{year}{2021}).

\bibitem[{\citenamefont{Liu et~al.}(2021)\citenamefont{Liu, He, and
  Cao}}]{PhysRevD.103.043005}
\bibinfo{author}{\bibfnamefont{X.}~\bibnamefont{Liu}},
  \bibinfo{author}{\bibfnamefont{X.}~\bibnamefont{He}}, \bibnamefont{and}
  \bibinfo{author}{\bibfnamefont{Z.}~\bibnamefont{Cao}},
  \bibinfo{journal}{Phys. Rev. D} \textbf{\bibinfo{volume}{103}},
  \bibinfo{pages}{043005} (\bibinfo{year}{2021}).

\bibitem[{\citenamefont{Bondi et~al.}(1962)\citenamefont{Bondi, Van~der Burg,
  and Metzner}}]{BonVanMet62}
\bibinfo{author}{\bibfnamefont{H.}~\bibnamefont{Bondi}},
  \bibinfo{author}{\bibfnamefont{M.}~\bibnamefont{Van~der Burg}},
  \bibnamefont{and} \bibinfo{author}{\bibfnamefont{A.}~\bibnamefont{Metzner}},
  \bibinfo{journal}{Proceedings of the Royal Society of London. Series A.
  Mathematical and Physical Sciences} \textbf{\bibinfo{volume}{269}},
  \bibinfo{pages}{21} (\bibinfo{year}{1962}).

\bibitem[{\citenamefont{Sachs}(1962)}]{Sac62}
\bibinfo{author}{\bibfnamefont{R.~K.} \bibnamefont{Sachs}},
  \bibinfo{journal}{Proceedings of the Royal Society of London. Series A.
  Mathematical and Physical Sciences} \textbf{\bibinfo{volume}{270}},
  \bibinfo{pages}{103} (\bibinfo{year}{1962}).

\bibitem[{\citenamefont{Penrose and Rindler}(1988)}]{PenRin88}
\bibinfo{author}{\bibfnamefont{R.}~\bibnamefont{Penrose}} \bibnamefont{and}
  \bibinfo{author}{\bibfnamefont{W.}~\bibnamefont{Rindler}},
  \emph{\bibinfo{title}{Spinors and space-time: Volume 1 and Volume 2}}
  (\bibinfo{publisher}{Cambridge University Press}, \bibinfo{year}{1988}).

\bibitem[{\citenamefont{He and Cao}(2015)}]{he2015new}
\bibinfo{author}{\bibfnamefont{X.}~\bibnamefont{He}} \bibnamefont{and}
  \bibinfo{author}{\bibfnamefont{Z.}~\bibnamefont{Cao}},
  \bibinfo{journal}{International Journal of Modern Physics D}
  \textbf{\bibinfo{volume}{24}}, \bibinfo{pages}{1550081}
  (\bibinfo{year}{2015}).

\bibitem[{\citenamefont{He et~al.}(2016)\citenamefont{He, Cao, and
  Jing}}]{he2016asymptotical}
\bibinfo{author}{\bibfnamefont{X.}~\bibnamefont{He}},
  \bibinfo{author}{\bibfnamefont{Z.}~\bibnamefont{Cao}}, \bibnamefont{and}
  \bibinfo{author}{\bibfnamefont{J.}~\bibnamefont{Jing}},
  \bibinfo{journal}{International Journal of Modern Physics D}
  \textbf{\bibinfo{volume}{25}}, \bibinfo{pages}{1650086}
  (\bibinfo{year}{2016}).

\bibitem[{\citenamefont{Sun et~al.}(2019)\citenamefont{Sun, Cao, and
  He}}]{sun2019binary}
\bibinfo{author}{\bibfnamefont{B.}~\bibnamefont{Sun}},
  \bibinfo{author}{\bibfnamefont{Z.}~\bibnamefont{Cao}}, \bibnamefont{and}
  \bibinfo{author}{\bibfnamefont{X.}~\bibnamefont{He}},
  \bibinfo{journal}{SCIENCE CHINA Physics, Mechanics \& Astronomy}
  \textbf{\bibinfo{volume}{62}}, \bibinfo{pages}{40421} (\bibinfo{year}{2019}).

\bibitem[{\citenamefont{Buonanno et~al.}(2007)\citenamefont{Buonanno, Cook, and
  Pretorius}}]{PhysRevD.75.124018}
\bibinfo{author}{\bibfnamefont{A.}~\bibnamefont{Buonanno}},
  \bibinfo{author}{\bibfnamefont{G.~B.} \bibnamefont{Cook}}, \bibnamefont{and}
  \bibinfo{author}{\bibfnamefont{F.}~\bibnamefont{Pretorius}},
  \bibinfo{journal}{Phys. Rev. D} \textbf{\bibinfo{volume}{75}},
  \bibinfo{pages}{124018} (\bibinfo{year}{2007}).

\bibitem[{\citenamefont{Br\"ugmann et~al.}(2008)\citenamefont{Br\"ugmann,
  Gonz\'alez, Hannam, Husa, Sperhake, and Tichy}}]{PhysRevD.77.024027}
\bibinfo{author}{\bibfnamefont{B.}~\bibnamefont{Br\"ugmann}},
  \bibinfo{author}{\bibfnamefont{J.~A.} \bibnamefont{Gonz\'alez}},
  \bibinfo{author}{\bibfnamefont{M.}~\bibnamefont{Hannam}},
  \bibinfo{author}{\bibfnamefont{S.}~\bibnamefont{Husa}},
  \bibinfo{author}{\bibfnamefont{U.}~\bibnamefont{Sperhake}}, \bibnamefont{and}
  \bibinfo{author}{\bibfnamefont{W.}~\bibnamefont{Tichy}},
  \bibinfo{journal}{Phys. Rev. D} \textbf{\bibinfo{volume}{77}},
  \bibinfo{pages}{024027} (\bibinfo{year}{2008}).

\bibitem[{\citenamefont{Cao et~al.}(2008)\citenamefont{Cao, Yo, and
  Yu}}]{PhysRevD.78.124011}
\bibinfo{author}{\bibfnamefont{Z.}~\bibnamefont{Cao}},
  \bibinfo{author}{\bibfnamefont{H.-J.} \bibnamefont{Yo}}, \bibnamefont{and}
  \bibinfo{author}{\bibfnamefont{J.-P.} \bibnamefont{Yu}},
  \bibinfo{journal}{Phys. Rev. D} \textbf{\bibinfo{volume}{78}},
  \bibinfo{pages}{124011} (\bibinfo{year}{2008}).

\bibitem[{\citenamefont{Favata}(2009{\natexlab{a}})}]{Fav09a}
\bibinfo{author}{\bibfnamefont{M.}~\bibnamefont{Favata}},
  \bibinfo{journal}{Phys. Rev. D} \textbf{\bibinfo{volume}{80}},
  \bibinfo{pages}{024002} (\bibinfo{year}{2009}{\natexlab{a}}).

\bibitem[{\citenamefont{Ashtekar et~al.}(2020)\citenamefont{Ashtekar,
  De~Lorenzo, and Khera}}]{Ashtekar:2019viz}
\bibinfo{author}{\bibfnamefont{A.}~\bibnamefont{Ashtekar}},
  \bibinfo{author}{\bibfnamefont{T.}~\bibnamefont{De~Lorenzo}},
  \bibnamefont{and} \bibinfo{author}{\bibfnamefont{N.}~\bibnamefont{Khera}},
  \bibinfo{journal}{General Relativity and Gravitation}
  \textbf{\bibinfo{volume}{52}}, \bibinfo{pages}{107} (\bibinfo{year}{2020}).

\bibitem[{\citenamefont{Mitman et~al.}(2021)\citenamefont{Mitman, Iozzo, Khera,
  Boyle, De~Lorenzo, Deppe, Kidder, Moxon, Pfeiffer, Scheel
  et~al.}}]{PhysRevD.103.024031}
\bibinfo{author}{\bibfnamefont{K.}~\bibnamefont{Mitman}},
  \bibinfo{author}{\bibfnamefont{D.~A.~B.} \bibnamefont{Iozzo}},
  \bibinfo{author}{\bibfnamefont{N.}~\bibnamefont{Khera}},
  \bibinfo{author}{\bibfnamefont{M.}~\bibnamefont{Boyle}},
  \bibinfo{author}{\bibfnamefont{T.}~\bibnamefont{De~Lorenzo}},
  \bibinfo{author}{\bibfnamefont{N.}~\bibnamefont{Deppe}},
  \bibinfo{author}{\bibfnamefont{L.~E.} \bibnamefont{Kidder}},
  \bibinfo{author}{\bibfnamefont{J.}~\bibnamefont{Moxon}},
  \bibinfo{author}{\bibfnamefont{H.~P.} \bibnamefont{Pfeiffer}},
  \bibinfo{author}{\bibfnamefont{M.~A.} \bibnamefont{Scheel}},
  \bibnamefont{et~al.}, \bibinfo{journal}{Phys. Rev. D}
  \textbf{\bibinfo{volume}{103}}, \bibinfo{pages}{024031}
  (\bibinfo{year}{2021}).

\bibitem[{\citenamefont{Pollney and Reisswig}(2011)}]{PolRei11}
\bibinfo{author}{\bibfnamefont{D.}~\bibnamefont{Pollney}} \bibnamefont{and}
  \bibinfo{author}{\bibfnamefont{C.}~\bibnamefont{Reisswig}},
  \bibinfo{journal}{The Astrophysical Journal Letters}
  \textbf{\bibinfo{volume}{732}}, \bibinfo{pages}{L13} (\bibinfo{year}{2011}).

\bibitem[{\citenamefont{Favata}(2009{\natexlab{b}})}]{Fav09b}
\bibinfo{author}{\bibfnamefont{M.}~\bibnamefont{Favata}}, \bibinfo{journal}{The
  Astrophysical Journal Letters} \textbf{\bibinfo{volume}{696}},
  \bibinfo{pages}{L159} (\bibinfo{year}{2009}{\natexlab{b}}).

\bibitem[{\citenamefont{Varma et~al.}(2019{\natexlab{a}})\citenamefont{Varma,
  Gerosa, Stein, H\'ebert, and Zhang}}]{PhysRevLett.122.011101}
\bibinfo{author}{\bibfnamefont{V.}~\bibnamefont{Varma}},
  \bibinfo{author}{\bibfnamefont{D.}~\bibnamefont{Gerosa}},
  \bibinfo{author}{\bibfnamefont{L.~C.} \bibnamefont{Stein}},
  \bibinfo{author}{\bibfnamefont{F.~m.~c.} \bibnamefont{H\'ebert}},
  \bibnamefont{and} \bibinfo{author}{\bibfnamefont{H.}~\bibnamefont{Zhang}},
  \bibinfo{journal}{Phys. Rev. Lett.} \textbf{\bibinfo{volume}{122}},
  \bibinfo{pages}{011101} (\bibinfo{year}{2019}{\natexlab{a}}).

\bibitem[{\citenamefont{Varma et~al.}(2019{\natexlab{b}})\citenamefont{Varma,
  Field, Scheel, Blackman, Gerosa, Stein, Kidder, and
  Pfeiffer}}]{PhysRevResearch.1.033015}
\bibinfo{author}{\bibfnamefont{V.}~\bibnamefont{Varma}},
  \bibinfo{author}{\bibfnamefont{S.~E.} \bibnamefont{Field}},
  \bibinfo{author}{\bibfnamefont{M.~A.} \bibnamefont{Scheel}},
  \bibinfo{author}{\bibfnamefont{J.}~\bibnamefont{Blackman}},
  \bibinfo{author}{\bibfnamefont{D.}~\bibnamefont{Gerosa}},
  \bibinfo{author}{\bibfnamefont{L.~C.} \bibnamefont{Stein}},
  \bibinfo{author}{\bibfnamefont{L.~E.} \bibnamefont{Kidder}},
  \bibnamefont{and} \bibinfo{author}{\bibfnamefont{H.~P.}
  \bibnamefont{Pfeiffer}}, \bibinfo{journal}{Phys. Rev. Research}
  \textbf{\bibinfo{volume}{1}}, \bibinfo{pages}{033015}
  (\bibinfo{year}{2019}{\natexlab{b}}).

\bibitem[{\citenamefont{Varma et~al.}(2020)\citenamefont{Varma, Isi, and
  Biscoveanu}}]{PhysRevLett.124.101104}
\bibinfo{author}{\bibfnamefont{V.}~\bibnamefont{Varma}},
  \bibinfo{author}{\bibfnamefont{M.}~\bibnamefont{Isi}}, \bibnamefont{and}
  \bibinfo{author}{\bibfnamefont{S.}~\bibnamefont{Biscoveanu}},
  \bibinfo{journal}{Phys. Rev. Lett.} \textbf{\bibinfo{volume}{124}},
  \bibinfo{pages}{101104} (\bibinfo{year}{2020}).

\bibitem[{\citenamefont{Abbott et~al.}(2020)}]{2020arXiv201014527A}
\bibinfo{author}{\bibfnamefont{R.}~\bibnamefont{Abbott}} \bibnamefont{et~al.},
  \bibinfo{journal}{arXiv e-prints} \bibinfo{eid}{arXiv:2010.14527}
  (\bibinfo{year}{2020}), \eprint{2010.14527}.

\bibitem[{\citenamefont{Abbott et~al.}(2019)}]{PhysRevX.9.031040}
\bibinfo{author}{\bibfnamefont{B.~P.} \bibnamefont{Abbott}}
  \bibnamefont{et~al.} (\bibinfo{collaboration}{LIGO Scientific Collaboration
  and Virgo Collaboration}), \bibinfo{journal}{Phys. Rev. X}
  \textbf{\bibinfo{volume}{9}}, \bibinfo{pages}{031040} (\bibinfo{year}{2019}).

\bibitem[{\citenamefont{Romero-Shaw et~al.}(2020)}]{10.1093.mnras.staa2850}
\bibinfo{author}{\bibfnamefont{I.~M.} \bibnamefont{Romero-Shaw}}
  \bibnamefont{et~al.}, \bibinfo{journal}{Monthly Notices of the Royal
  Astronomical Society} \textbf{\bibinfo{volume}{499}}, \bibinfo{pages}{3295}
  (\bibinfo{year}{2020}).

\bibitem[{\citenamefont{Lindblom and Cutler}(2016)}]{PhysRevD.94.124030}
\bibinfo{author}{\bibfnamefont{L.}~\bibnamefont{Lindblom}} \bibnamefont{and}
  \bibinfo{author}{\bibfnamefont{C.}~\bibnamefont{Cutler}},
  \bibinfo{journal}{Phys. Rev. D} \textbf{\bibinfo{volume}{94}},
  \bibinfo{pages}{124030} (\bibinfo{year}{2016}).

\bibitem[{\citenamefont{P\"urrer and Haster}(2020)}]{PhysRevResearch.2.023151}
\bibinfo{author}{\bibfnamefont{M.}~\bibnamefont{P\"urrer}} \bibnamefont{and}
  \bibinfo{author}{\bibfnamefont{C.-J.} \bibnamefont{Haster}},
  \bibinfo{journal}{Phys. Rev. Research} \textbf{\bibinfo{volume}{2}},
  \bibinfo{pages}{023151} (\bibinfo{year}{2020}).

\bibitem[{\citenamefont{Collaboration et~al.}(2020)\citenamefont{Collaboration,
  Collaboration et~al.}}]{ligo2020gwtc}
\bibinfo{author}{\bibfnamefont{L.~S.} \bibnamefont{Collaboration}},
  \bibinfo{author}{\bibfnamefont{V.}~\bibnamefont{Collaboration}},
  \bibnamefont{et~al.}, \bibinfo{journal}{LIGO-P2000223}
  (\bibinfo{year}{2020}).

\bibitem[{\citenamefont{Garc{\'\i}a-Quir{\'o}s
  et~al.}(2020)\citenamefont{Garc{\'\i}a-Quir{\'o}s, Colleoni, Husa,
  Estell{\'e}s, Pratten, Ramos-Buades, Mateu-Lucena, and
  Jaume}}]{garcia2020multimode}
\bibinfo{author}{\bibfnamefont{C.}~\bibnamefont{Garc{\'\i}a-Quir{\'o}s}},
  \bibinfo{author}{\bibfnamefont{M.}~\bibnamefont{Colleoni}},
  \bibinfo{author}{\bibfnamefont{S.}~\bibnamefont{Husa}},
  \bibinfo{author}{\bibfnamefont{H.}~\bibnamefont{Estell{\'e}s}},
  \bibinfo{author}{\bibfnamefont{G.}~\bibnamefont{Pratten}},
  \bibinfo{author}{\bibfnamefont{A.}~\bibnamefont{Ramos-Buades}},
  \bibinfo{author}{\bibfnamefont{M.}~\bibnamefont{Mateu-Lucena}},
  \bibnamefont{and} \bibinfo{author}{\bibfnamefont{R.}~\bibnamefont{Jaume}},
  \bibinfo{journal}{Physical Review D} \textbf{\bibinfo{volume}{102}},
  \bibinfo{pages}{064002} (\bibinfo{year}{2020}).

\bibitem[{\citenamefont{Talbot et~al.}(2018)\citenamefont{Talbot, Thrane,
  Lasky, and Lin}}]{PhysRevD.98.064031}
\bibinfo{author}{\bibfnamefont{C.}~\bibnamefont{Talbot}},
  \bibinfo{author}{\bibfnamefont{E.}~\bibnamefont{Thrane}},
  \bibinfo{author}{\bibfnamefont{P.~D.} \bibnamefont{Lasky}}, \bibnamefont{and}
  \bibinfo{author}{\bibfnamefont{F.}~\bibnamefont{Lin}},
  \bibinfo{journal}{Phys. Rev. D} \textbf{\bibinfo{volume}{98}},
  \bibinfo{pages}{064031} (\bibinfo{year}{2018}).

\end{thebibliography}

\end{document}